\documentclass[useAMS,usenatbib]{mn2e}

\usepackage{txfonts,graphicx,amssymb,subfigure}
\usepackage[normalem]{ulem}

\title{Magnetic fields in fully convective M-dwarfs: oscillatory dynamos vs bistability}
\author[L.L.~Kitchatonov et al.]
       {L.L.~Kitchatinov$^{1,2}$, D.~Moss$\,^{3}$ and  D.~Sokoloff$^4$\\
$^1$Institute of Solar-Terrestrial Physics, PO Box 291, Irkutsk 664033, Russia  \\
$^2$Pulkovo Astronomical Observatory, St. Petersburg, 196140, Russia\\
$^3$School of Mathematics, University of Manchester, Oxford Road,
Manchester, M13 9PL, UK\\
$^4$Department of Physics, Moscow University, 119899, Moscow, Russia
}

\date{Accepted  .... Received  ....; in original form }

\pagerange{\pageref{firstpage}--\pageref{lastpage}}
\pubyear{2014}
\begin{document}
\maketitle

\label{firstpage}

\begin{abstract}
M-dwarfs demonstrate two types of activity: 1) strong (kilogauss) almost axisymmetric poloidal magnetic fields; and
2) considerably weaker nonaxisymmetric fields, sometimes including a
substantial toroidal component. Dynamo bistability has been proposed as an explanation.
However it is not
straightforward to obtain such a bistability in dynamo models.
On the other hand, the solar magnetic dipole at times
of magnetic field inversion becomes transverse to the rotation axis,
while the magnetic field becomes weaker at
times far from that of  inversion. Thus the
 Sun resembles a star with the second type of activity. We suggest that M-dwarfs can
have magnetic cycles, and that M-dwarfs with the second type of activity can just
be stars observed at times of
magnetic field inversion. Then the relative number of M-dwarfs with the second type
of activity can be used in the framework of this model
to determine parameters of stellar convection near the surface.

\end{abstract}
\section{Introduction}

The  efforts of many observing teams over several decades have provided rich data
concerning magnetic activity in stars of various spectral classes.
However in many cases the general form of the activity of stars of a
given spectral class remains debatable. In particular, observations find two distinct magnetic topologies for M-dwarfs:
1) strong (kilogauss) almost axisymmetric poloidal magnetic fields, and
2) considerably weaker nonaxisymmetric fields, sometimes including a
substantial toroidal component (Morin et al. 2010).
Stars with different topologies can have roughly the same rotation
rates, mass, and other parameters. This finding has been interpreted as dynamo
bistability, i.e. as two different regimes of dynamo action being stable for
the same set of stellar parameters (Morin et al. 2011a,b; Gastine et al.  2013).
However, two stable regimes of a dynamo in the same parameter range seems not to be
very probable (cf. e.g. R\"adler et al. 1990; Moss \& Sokoloff 2009).
On the other hand,  Chabrier \& K\"uker
(2006) suggest that fully convective stars host a non-axisymmetric global dynamo
(such a possibility was also suggested
in a slightly different context by, e.g., Barker \& Moss 1994). This dynamo
is steady (equatorial dipole drifting in longitude) and therefore showing no reversals.
This model relies on the
observational fact (Barnes et al. 2005) supported by numerical simulations
(K\"uker \& R\"udiger 2005) that differential rotation in fully convective stars is
very small. As a result Chabrier \& K\"uker (2006) neglect differential rotation and
considered an $\alpha^2$ dynamo.

The efficiency of differential rotation in winding up
 magnetic fields can be estimated by the dimensionless dynamo number
\begin{equation}
    C_\Omega = \frac{\Delta\Omega H^2}{\eta_{_\mathrm{T}}},
    \label{C_omega}
\end{equation}
where $\Delta\Omega$ is the angular velocity variation within the convection zone, $H$ is the convection zone thickness and $\eta_{_\mathrm{T}}$ is the eddy magnetic diffusivity. Differential rotation modelling suggests that the decrease in $\Delta\Omega$ with decreasing temperature
can be compensated by a simultaneous decrease in $\eta_{_\mathrm{T}}$,
 so that $C_\Omega$ actually {\em increases} in cooler stars
(Kitchatinov \& Olemskoy 2011; Kitchatinov 2013). This means that the M-dwarfs have
small but very efficient differential rotation and
it is quite probable that they host oscillatory, axisymmetric mean-field dynamos.

Taking all this into account, it seems reasonable to consider another possible
explanation of the M-dwarf activity phenomenon.
We suggest that magnetic cycles in the form of oscillatory axisymmetric fields
 are present within the population of M-dwarfs,
and that the stars that are observed to have weak nonaxisymmetric fields are
observed at epochs of reversal,
with the strong axisymmetric dipoles being present at the analogues of solar maxima.
There are strong observational indications for a difference in magnetic topologies
between fully convective stars and stars with convective envelopes,
with magnetic fields of fully convective stars being dominated by axisymmetric poloidal
configurations (Gregory et al. 2012). Nevertheless, an analogy with the dynamics of
the (relatively weak) solar poloidal field is possible and can be useful.
Note that the solar activity is determined mainly by the toroidal magnetic field.
There is a time lag between the cyclic oscillations of toroidal and poloidal magnetic
fields so that the solar magnetic dipole is strong during the minima of solar activity.
A straightforward verification of this explanation could be performed by
monitoring of a sample of M-dwarfs over times exceeding the expected period.
Estimates in the next section suggest however that magnetic cycles in M-dwarfs can last
for several decades. If M-dwarfs do have cycles, the cycles could therefore be
substantially longer than that of the Sun, while at the moment only a 3-year monitoring
(2006-2009) is available (Morin et al. 2010).

A straightforward objection to the last explanation could be that solar mean-field
dynamo models driven by differential rotation and mirror-asymmetric turbulence
($\alpha \Omega$-dynamos) give axisymmetric mean magnetic fields, and the
magnetic dipole has to vanish during its inversion, and be parallel to the rotation axis
between its reversals. Recent progress in understanding solar observations and
the solar dynamo  (Moss et al. 2013a,b) have provided a new understanding
of the reversals of the solar magnetic dipole. This allows us to elaborate the idea
under discussion quantitatively, and to suggest a way to verify it that does
not ultimately require a long-term monitoring programme
(which of course still remains highly desirable).

\section{Oscillatory dynamos in M-dwarfs}

Taking into account the inherent uncertainties and arbitrariness
of dynamo modelling for wide classes of stars,
and that it is
far from clear how generic any particular model (set-up, choice of parameters, etc)
can be, nevertheless we performed
an exploratory modelling to establish the possibility of oscillatory dynamos
existing in M-dwarfs, as follows.

\begin{figure}
\begin{center}
\includegraphics[width= 7 cm]{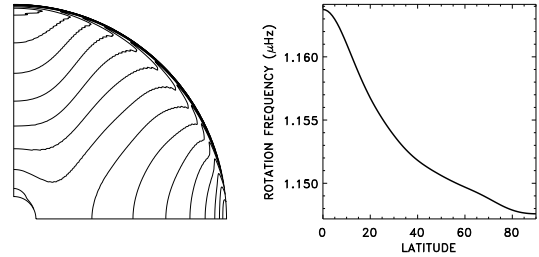}
\end{center}
\caption{\label{diff}
Isorotational curves (left) and surface 
rotation frequency versus latitude (right) for the reference M-dwarf model.}
\end{figure}

We computed the differential rotation for a star
with $M=0.3 M_\odot$, rotating with period of 10 days, using the numerical mean-field
model of Kitchatinov \& Olemskoy (2011). The result is shown in Fig.~\ref{diff}.
Such a  star is fully convective, but the model requires an (in this case artificial)
inner boundary of the convection zone to be  imposed at $r = 0.1R_\mathrm{star}$.
The model does not prescribe eddy transport coefficients but estimates them
from the entropy gradient, so that the $C_\Omega$ parameter of Eq.~(\ref{C_omega})
can also be estimated. Taking the turbulent Prandtl number,
\begin{equation}
    P_\mathrm{m} = \nu_{_\mathrm{T}}/\eta_{_\mathrm{T}},
    \label{P_m}
\end{equation}
to have the value unity gives $C_\Omega \simeq 290$, which is not small
in spite of the small differential rotation in Fig.~\ref{diff}.
The eddy viscosity $\nu_{_\mathrm{T}} \simeq 1.2\times 10^{11}$\,cm$^2$s$^{-1}$
used in this estimate is
taken at the middle radius $r = R_\mathrm{star}/2$ ($R_\mathrm{star} = 212$~Mm).
Thus the diffusive time is $\sim100$ years. If the cycle time is close to the
diffusion time, as it is for the Sun, very long cycles can be expected.
The circulation time for the meridional flow is also long, $\sim$60 yr
(the typical flow velocity is 10 cm/s except in the thin surface boundary layer,
where it is about 4 m/s). Thus the cycle time for an advection-dominated dynamo will
 also be long. We use these estimates to obtain the reference values for the dynamo
governing parameters for M-dwarfs and experiment with numbers around these
reference quantities.

\begin{figure}
\begin{center}
\includegraphics[width= 3.5 cm]{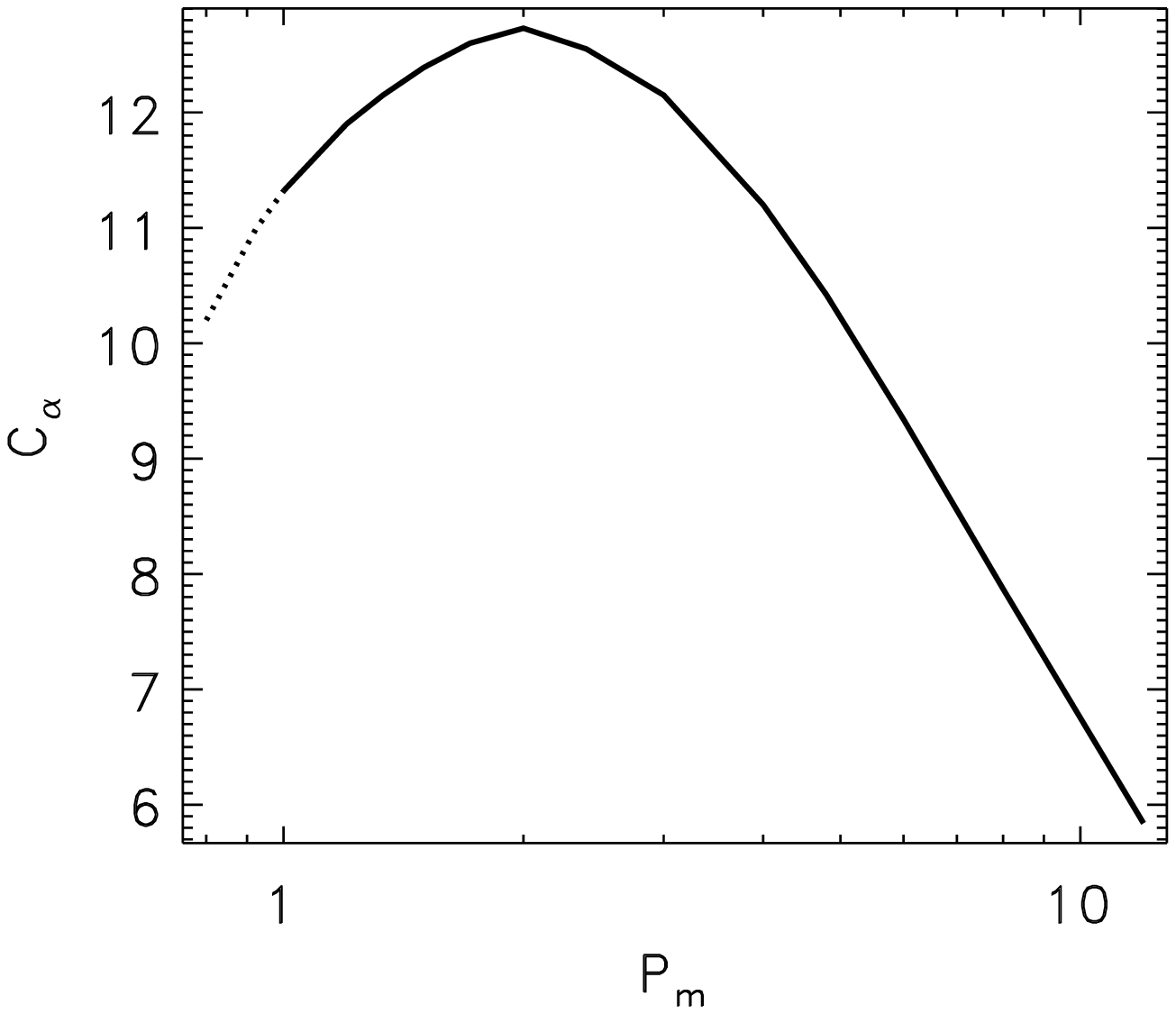}
\hspace{0.1truecm}
\includegraphics[width= 3.6 cm]{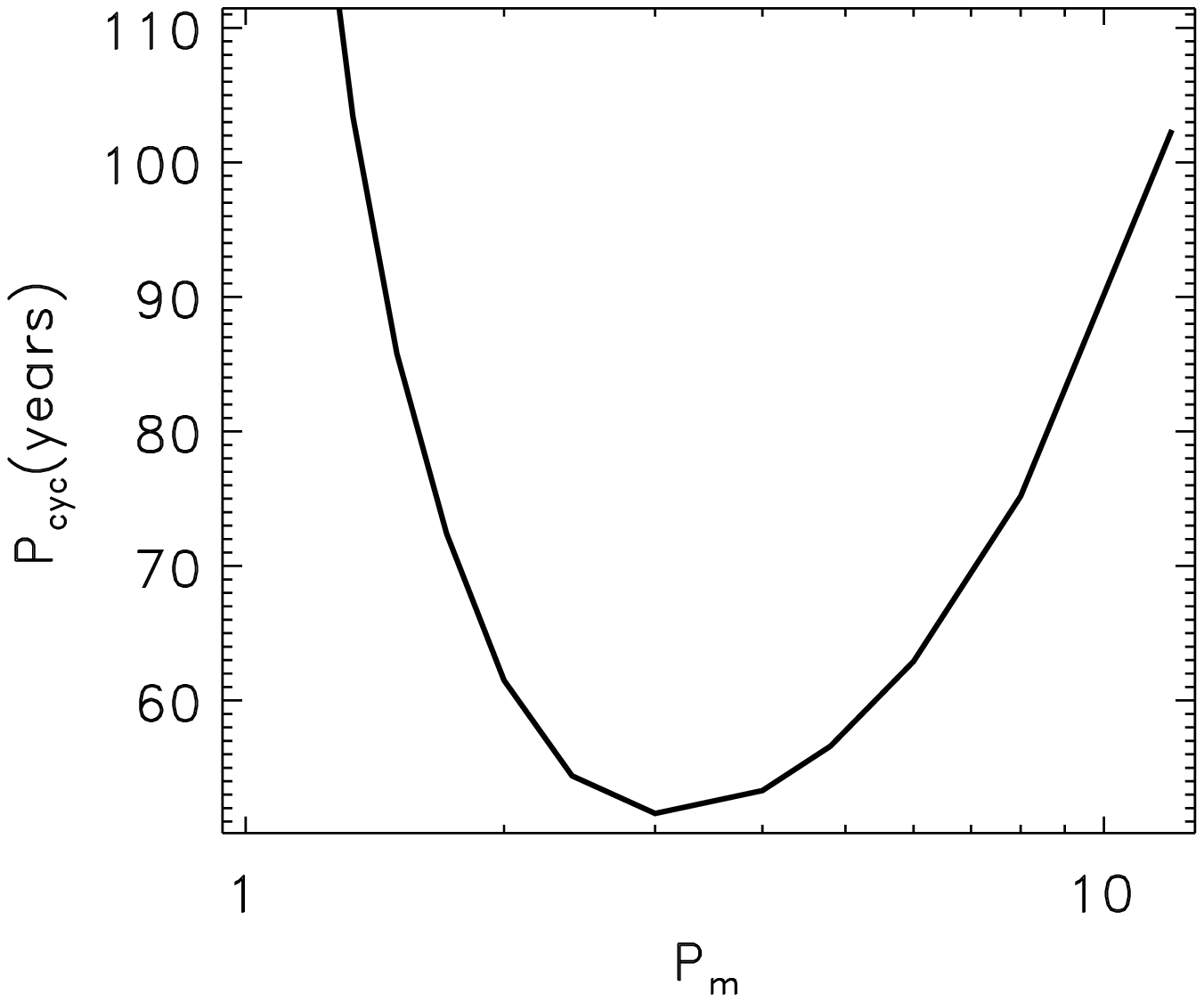}
\end{center}
\caption{\label{dwarfdyn}
Dynamo model for a M-dwarf: left - marginal values of $C_\alpha$ (\ref{C_alpha}) 
for axisymmetric dynamo modes of dipolar parity, 
dots indicate steady modes  and the solid line shows the oscillatory modes,
right - the corresponding oscillation periods.}
\end{figure}
Fig.~\ref{dwarfdyn} demonstrates the results for the simplest dynamo model with
uniform diffusion, $\alpha = \alpha_0\cos\theta$ uniform with radius and $\cos\theta$ dependence
on co-latitude, isotropic diffusion and alpha. 
The model estimates the threshold value of the dimensionless parameter,
\begin{equation}
    C_\alpha = \frac{\alpha_0 R_\mathrm{star}}{\eta_{_\mathrm{T}}} ,
    \label{C_alpha}
\end{equation}
for onset of dynamo together with the field structure and oscillation frequency. 
The eddy viscosity is estimated by
the differential rotation code.
We take the magnetic Prandtl number of Eq.~(\ref{P_m}) as a free parameter.
We conclude from this limited modelling that an oscillatory dynamo
model with solar type behaviour can be considered as a viable option at least at
the present level of investigation. We stress that further modelling of
dynamo action in M-dwarfs remains highly desirable.

The transition from steady to oscillatory dynamos with increasing $P_\mathrm{m}$,
seen in Fig.~\ref{dwarfdyn}, indicates that the dynamo operates in the $\alpha^2\Omega$ regime.
In this regime, poloidal fields can be expected to be much stronger than in solar-type
$\alpha\Omega$ dynamos. Equipartition poloidal fields of kilogauss strength
$B\sim \sqrt{4\pi\rho}\ u_\mathrm{conv} \sim 10^3$\,G can be expected for the
fully convective M-dwarfs ($\rho = 37$~g\,cm$^{-3}$,
$u_\mathrm{conv} = 50$~cm\,s$^{-1}$ at $r = R_\mathrm{star}/2$ from 
the model of the $0.3M_\odot$ star structure
).

\section{Reversals of the solar dipole as a paradigm for reversals in M-dwarfs}

We start by noting that many solar observers (e.g.  Antonucci 1974; Zhukov \& Veselovsky 2000;
Livshits \& Obridko 2006) have reported that the solar magnetic dipole
does not vanish during the reversal. Recently De Rosa
et al. (2012) presented a comprehensive data sample for the two last solar activity cycles
which convincingly demonstrate that this is indeed the case.

Moss et al. (2013a) suggested how to resolve this apparent contradiction between expectations
from dynamo modelling and observation.
The point is that a mean-field dynamo model deals with {\it mean} magnetic field and
the averaging is performed over an ensemble of convective velocity cells,
 while the observational magnetic dipole data refer to
{\it large-scale} magnetic field. Both quantities coincide for an
infinitely large ensemble of convective cells, but in practice the number of cells is
only moderately large ($N=10^4$ is a crude relevant
estimate,  (see Moss et al. 2013a for details). Because the convective cell
ensemble contains a not extremely large number of cells,
large-scale fluctuations of magnetic field arise which yield a
fluctuating component $\delta d$ of the solar magnetic dipole $d$,  of order

\begin{equation}
\delta d/d \propto (b/B) N^{-1/2} (B_P/B_T) \, .
\label{estim}
\end{equation}
Here $b$ is the r.m.s. value of small-scale magnetic field, i.e. the
magnetic fluctuations, $B$ is the typical value of
the mean magnetic field which is determined mainly by the toroidal magnetic field $B_T$,
and the factor $B_T/B_P$ takes
into account that the magnetic dipole moment is determined by the poloidal magnetic field $B_P$.  Moss et al. (2013a)
demonstrated that the estimate for the solar magnetic dipole at the epoch of inversion,
 based on Eq.~(\ref{estim}) and
the available information about the relevant solar parameters is
consistent with this picture.

\begin{figure}
\begin{center}
\includegraphics[width= 7 cm]{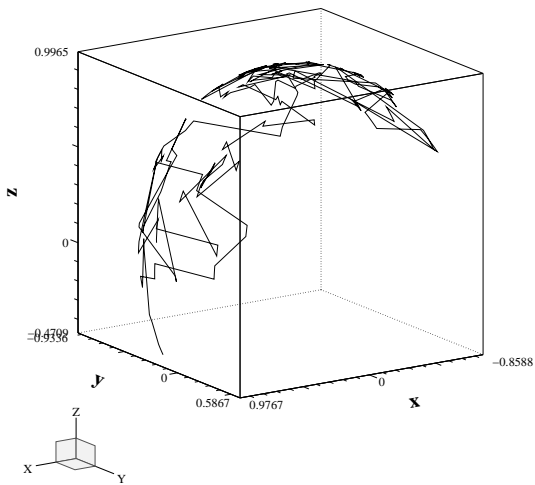}
\end{center}
\caption{\label{migr}
Trajectory of the solar dipole during a reversal of the solar magnetic field}
\end{figure}

Moss et al. (2013b) compared this scenario with observational data more systematically,
 to learn that the interval during which
the fluctuating part of the magnetic dipole is larger than the part determined by
the mean magnetic field is about 4
months, i.e. about 3\% of the solar magnetic cycle.  Fig.~\ref{migr} shows (from an example
from the Moss et al. 2013b analysis) the migration of the dipole from one solar pole
to the other, and clearly demonstrates the random nature of the reversal.

\section{Rescaling the solar example}

Assuming that the magnetic activity of M-dwarfs is more-or-less similar to that
of the Sun,  we can use
Eq.~(\ref{estim}) to determine the relative time $\delta t/T$ during the magnetic activity
cycle during which the magnetic dipole is determined
by magnetic fluctuations and is strongly inclined to the rotation axis.
Because the scenario assumes that at that time the
axisymmetric dipolar field is minimal
we conclude that observations taken at that epoch would identify this
star as exhibiting the second type of activity.

If we consider a random sample of M-dwarfs then the value $\delta t/T$ immediately gives the relative number
$\delta n/N$ of M-dwarfs in the sample which demonstrates the second type of activity.

The temporal evolution of the axisymmetric solar magnetic dipole is approximately sinusoidal
(Moss et al. 2013a), i.e. much simpler than
the evolution of the mean sunspot number. Assuming that the same is valid for M-dwarfs,
and taking into account Eq.~(\ref{estim}) we can say what physical parameters
are required , e.g. $\delta n/N$,
to get a satisfactory correspondence with the observational data.
In particular, if we want
to explain that about 30\% of M-dwarfs demonstrate the second type of activity,
we have to assume that the number
of convective cells at the surface of M-dwarfs is about two order of magnitudes
lower than near the surface of the Sun, i.e. $10^2$ instead of
$10^4$. (Given the much greater relative depth of the convection zone in M-dwarfs compared
to the Sun, an increase in the size of convection cells is not implausible.) Then, Eq.~(\ref{estim}) reads that
$\delta d/d$ increases by  factor $10$
in comparison with the Sun
and because the oscillations of the dipole strength are near to sinusoidal,
the time at which fluctuations are
weaker than the mean value of the dipole grows by a factor $10^2$,
and now becomes about 1/3 of the cycle duration. As a result,
the relative number of M-dwarfs which demonstrate the second type of activity
increases by  a factor $10$,  giving 30\%.

\section{Discussion and conclusions}
\label{concl}

We have suggested a scenario which explains why observations identify
two types of magnetic activity in M-dwarfs.
We propose that a M-dwarf can either be observed at a time far from that of
 magnetic field inversion (giving the first type of activity), or
near the inversion (giving the second type).

A quantification of the scenario under discussion can be performed on the basis of
Eq.~(\ref{estim},) by determination of
the percentage of M-dwarfs that demonstrate the second type of activity.
We stress that such quantification does not
require an effort-consuming long-term monitoring of M-dwarf activity.
Of course, such a monitoring remains highly desirable for
further understanding of stellar magnetic activity.

Even a crude estimate of the relative number of M-dwarfs
with the second type of activity will provide in our framework a perspective
for monitoring activity in M-dwarfs. If, say, this number were about 10\%,
then it would only be necessary to perform monitoring during 1/10 of a stellar cycle
in order to observe a transition from one activity type to the other one in
a given object. Another problem of interest would be to determine how
this relative number varies between solar type stars and M-dwarfs.
Indeed, any star with periodic dynamo driven large-scale fields might
be expected to display similar behaviour, but whether this could
be detected by current observational techniques
is another matter.

\section*{Acknowledgments}
DS and LLK are grateful to the Russian Foundation for Basic Research for financial support under grants 12-02-00170 and 13-02-00277 respectively.

\end{document}